\documentclass[12pt,preprint]{aastex}

\slugcomment{ApJ Letters, Accepted Sep.~16, 2008}

\shorttitle{A Circumbinary Photoevaporated Disk in ONC}
\shortauthors{Robberto et al.}

\begin{document}

\title{Evidence for a Photoevaporated Circumbinary Disk in Orion\altaffilmark{1}}

\altaffiltext{1}{Based on observations made with the NASA/ESA Hubble Space Telescope, obtained at the Space Telescope Science Institute, which is operated by the Association of Universities for Research in Astronomy, Inc., under NASA contract NAS 5-26555. These observations are associated with program 10246.}

\author{M. Robberto, L. Ricci\altaffilmark{2}, N. Da Rio\altaffilmark{3} and D. R. Soderblom}
\affil{Space Telescope Science Institute, Baltimore, MD 21218}
\email{robberto@stsci.edu; lricci@eso.org; dario@mpia-hd.mpg.de; drs@stsci.edu}

\altaffiltext{2}{present address: European Southern Observatory, Garching bei M\"unchen, Germany}
\altaffiltext{3}{present address: Max-Planck-Institut f\"ur Astronomie, Heidelberg, Germany}

\begin{abstract}
We have found a photoevaporated disk in the Orion Nebula that includes a wide binary. HST/ACS observations of the proplyd 124-132 show two point-like sources separated by $0\farcs15$, or about 60 AU at the distance of Orion.  The two sources have nearly identical $I$ and $z$ magnitudes. We analyze the brightest component, Source~N, comparing the observed magnitudes with those predicted using a 1\,Myr Baraffe/NEXTGEN isochrone with different accretion luminosities and extinctions. We find that a low mass ($\simeq\!0.04~M_\odot$) brown dwarf $\sim$1~Myr old with mass accretion rate $\log\dot{ M}\simeq-10.3$,  typical for objects of this mass, and about 2~magnitudes of visual extinction provides the best fit to the data.  This is the first observation of a circumbinary disk undergoing photoevaporation and, if confirmed by spectroscopic observations, the first direct detection of a wide substellar pair still accreting and enshrouded in its circumbinary disk. 
\end{abstract}

\keywords{ISM: individual (Orion Nebula, M42) --- planetary systems: protoplanetary disks --- binaries: general --- stars: low-mass, brown dwarfs }

\section{Introduction and Observations}

We report on the discovery of a circumbinary disk seen in silhouette against the bright nebular background of the Orion Nebula. Using multicolor observations taken with the Hubble Space Telescope as part of the HST Treasury Program on the Orion Nebula Cluster, we show that source 124-132, a photoevaporated disk $\approx\!1\farcm5$ north of the Trapezium previously imaged in H$\alpha$ \citep{Smith+05}, harbors a binary.

Binaries represent a typical product of the gravitational collapse of cores with high angular momentum. In fact, the  majority of stars in star-forming regions are in binary or multiple systems (see Duchene 1999; Monin et~al.\ 2007; and for a review Mathieu 2000 and references therein). Also the ONC contains a large number of binary stars
\citep{Prosser+94, Padgett+97, 1998ApJ...500..825P, Koehler+06,Reipurth+07}, although the current counts indicate that the binary frequency, especially at the low-mass end, is lower by a factor 2 to 5 than in star-forming T~Associations like
Taurus-Auriga \citep{Koehler+06, Reipurth+07}.  Whether the relative paucity of binary systems in the ONC is due to the initial conditions of the cloud or to a ``feedback" effect, like the ejection of binary companions in the cluster core caused by close dynamic encounters, is still a subject of debate.

There can be three disks in a young binary system: two circumstellar disks and a circumbinary one
\citep{1993prpl.conf..749L, 1994ApJ...421..651A, 1997MNRAS.285...33B},
with a complex geometry due to the potential variety of alignments between the disks and orbital planes \citep{Monin+07}, and interacting through transfer of energy and angular momentum \citep{Gunther+02}. Disks have been detected around many spectroscopic binaries. Direct imaging of disks around wide, well separated binary systems, either circumstellar or circumbinary, is much rarer, the two most notable examples being the circumbinary disks around GG~Tau \citep{1994A&A...286..149D} and UY~Aur \citep{1998A&A...332..867D}.

The data presented here are part of the large dataset (520 images) of ACS exposures obtained for the \textit{HST Treasury Program on the Orion Nebula Cluster} (GO-10246). We mapped with ACS/WFC an area of about 450 square arcmin nearly centered on the Trapezium Cluster in five filters: F435W (Johnson~$B$, 420s); F555W (Johnson~$V$, 385s); F685N (H$\alpha$+[\ion{N}{2}]  $\lambda 6583$, 340s); F775W (Cousins~$I_\mathrm{C}$, 385s); and F850LP ($z$-band, 385s). Due to the adopted dithering strategy, most of the field has been
exposed two times so the total integration time is typically twice that listed.   The drizzled ACS images have been visually inspected and a master catalog of 219 circumstellar disks has been compiled \citep{Ricci+08}. Figure~1 is extracted from this last paper, which also provides  the fits files in electronic form. Further details on the \textit{HST Treasury Program} observing strategy and data products will be given elsewhere  (Robberto et~al.\ 2008, in preparation).

\section{Results}

\begin{figure}
\epsscale{.7}
\plotone{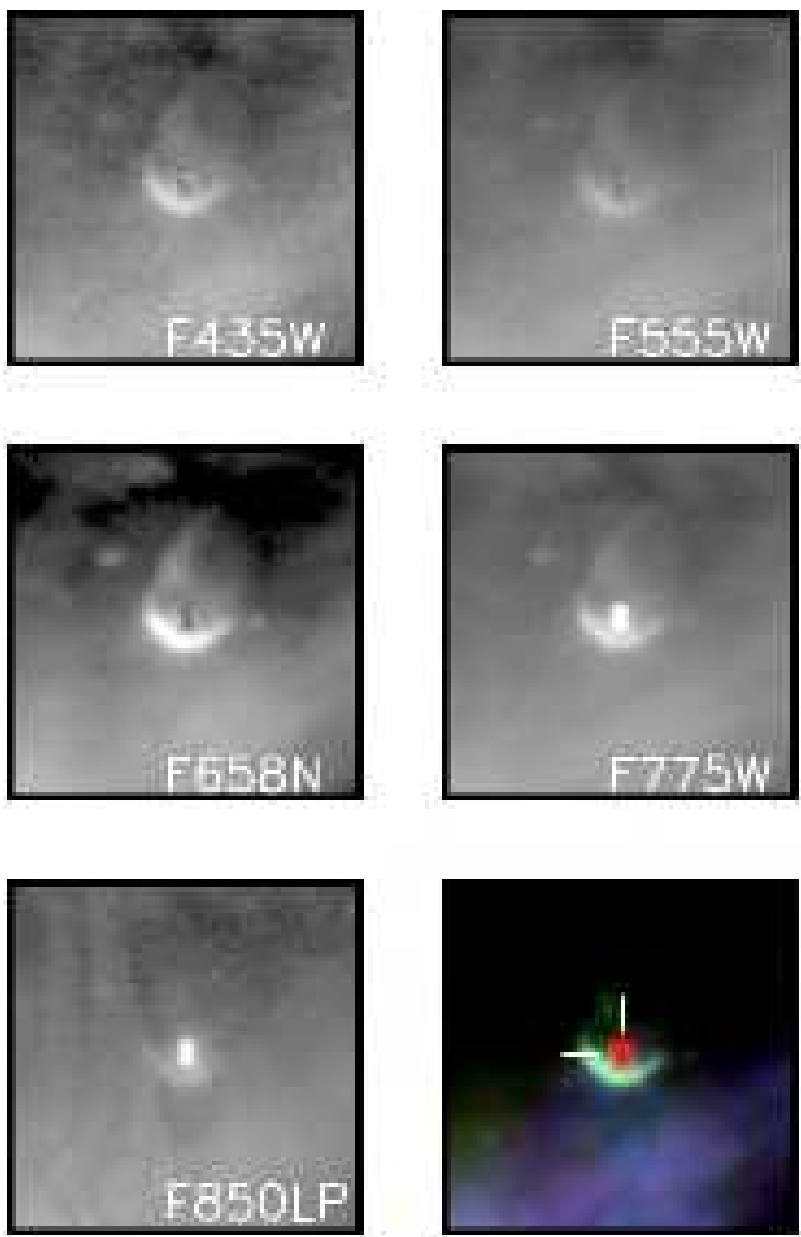}
 \caption{ACS/WFC images of 124-132; Each frame shows $100 \times 100$ ACS/WFC pixels, corresponding to $\sim\!\!5'' \times 5''$, or $\sim\!\!2000 \times 2000$~AU at the distance of the Orion Nebula, here assumed to be 414~pc \citep{2007A&A...474..515M}. The color picture at the bottom-right is a composite of the five filters with colors assigned in this way: the intensity of blue is given by the average between the fluxes measured in the F435W and F555W bands, the intensity of red by the average in the F775W and F850LP bands, and the intensity of green by the flux measured in the F658N filter only. Images
are oriented with north up and east to the left. \label{fig1}}
\end{figure}

Figure \ref{fig1} shows ACS/WFC images of source  124-132. \citet{Smith+05} have already illustrated the H$\alpha$ morphology of this object, a bright proplyd with a well resolved dark disk nearly perpendicular to the vertex of the ionization front. From their H$\alpha$ analysis, they estimate a disk tilt angle of $\simeq\!75^\circ$. The disk shows some point-like H$\alpha$\ emission on the west (right) side, which they interpret as either emission from a bipolar reflection nebula or a microjet perpendicular to the major disk axis.

A more careful look at the H$\alpha$ image shows that the point-like source is not centered on the dark disk axis,  but shifted approximately 2~pixels ($0\farcs1$) to the north, away from the ionization front. Since the brightness of the ionization front does not allow us to trace the full extension of the disk to the south, this asymmetry is probably real. The situation becomes clearer when our other images are considered. At shorter wavelengths (F435W and F555W), the point-like source remains visible to the west of the disk axes, suggesting that the silhouette disk is slightly tilted with the western face toward us. More interestingly, at longer wavelengths (F775W and F850LP), two point-like sources appear well resolved along the disk axes.

\begin{figure}
\epsscale{.80}
\plotone{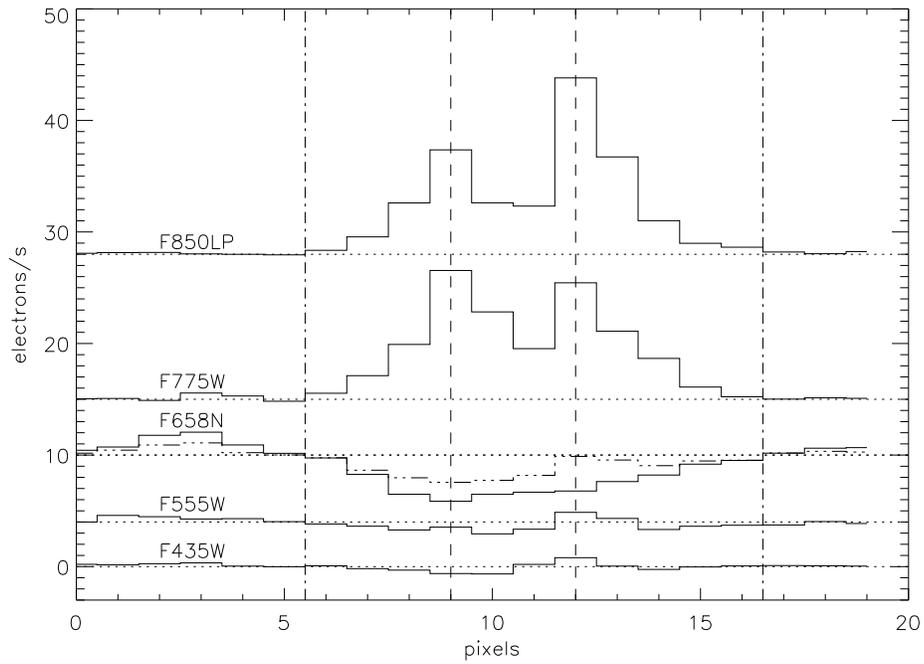}
\caption{North-south cuts (north is to the right) through the diameter of the silhouette disks for the 5 ACS filters (solid lines). 1 pixel (50 mas) corresponds to 21~AU at a distance of 420~pc. An offset has been added to the counts in each filter for illustrative purposes, with the horizontal dotted lines corresponding to the offset values. The horizontal dash-dot line for the F658N filter is the  adjacent column to the west. The position of the sources is marked by the vertical dashed lines, whereas the limits of the disk are indicated with vertical dash-dot lines. \label{fig2}}
\end{figure}

In Figure~\ref{fig2} we plot the counts measured through the central part of the disk (column~50 of the images in Fig.~\ref{fig1}). In order to remove the
contribution of the sky background and of the proplyd we have subtracted from each cut the counts measured two pixels to the west (column~52). Figure~\ref{fig2} clearly shows the two well resolved point sources appearing at the same position in the F775W and F850LP filters. The F658N cut reveals the extent of the disk and shows that the disk is darker and possibly wider at the location of Source~S. For the F658N filter we have also added a line, corresponding to column~51, to show the location of the H$\alpha$ source discussed by \citet{Smith+05}. Source~N remains visible in the F555W and F435W filters, whereas Source~S is perhaps barely detectable only in the F555W filter. In  Figure~\ref{fig2} we also indicate the limits of the disk, showing that the disk extends further on the northern side. This may  be due to the presence of the ionized rim, which prevents tracking the disk in the vicinity of the ionization front. In any case, we estimate a  disk diameter of approximately 10~pixels, i.e., 0\farcs5 ($\simeq\!200$~AU), whereas the projected separation of the two stars is $\sim$3~pixels, corresponding to 60~AU.

In Table \ref{tbl-2} we present the magnitudes of the two sources measured by integrating the counts in a 3~pixel diameter aperture; the sky has been measured at a close offset position free from ionized emission. We have applied
the ACS zero points (Vega photometric system) and aperture corrections from \citet{Sirianni+05}, averaging the encircled energy reported in Table~3 of Sirianni et~al.\ for aperture radii of 1 and 2~pixels. Being HST diffraction limited, the PSF in the widest filters depends on the source color. Following \citet{Sirianni+05}, we adjust the aperture correction in the F850LP filter to account for the extremely red colors of our sources. By repeating the measures with subpixel shifts of the centroid, we estimate an uncertainty of approximately 0.1~mag in the F775W and F850LP passbands, and 0.2~mag in the bluer filters due to  the faintness of the source and the highly non-uniform background. The most striking result is the similarity between the magnitudes and $I-z$ colors of the two sources, with both of them having F775W$-$F850LP~= 1.5.  

\section{Discussion}

Our images reveal the presence of a dark silhouette disk harboring two faint, similar  objects.  In the F775W and F850LP filters the sources appear clearly  point-like, well separated, and without any evidence of bipolar extended  emission from either sides of the disk, as normally observed when edge-on disks obscure their central stars \citep[][and references therein]{Luhman+07}. We will thus assume that we are seeing the reddened central stars rather than their diffuse  light scattered by the disk surface. 

\begin{figure}
\epsscale{.80}
\plotone{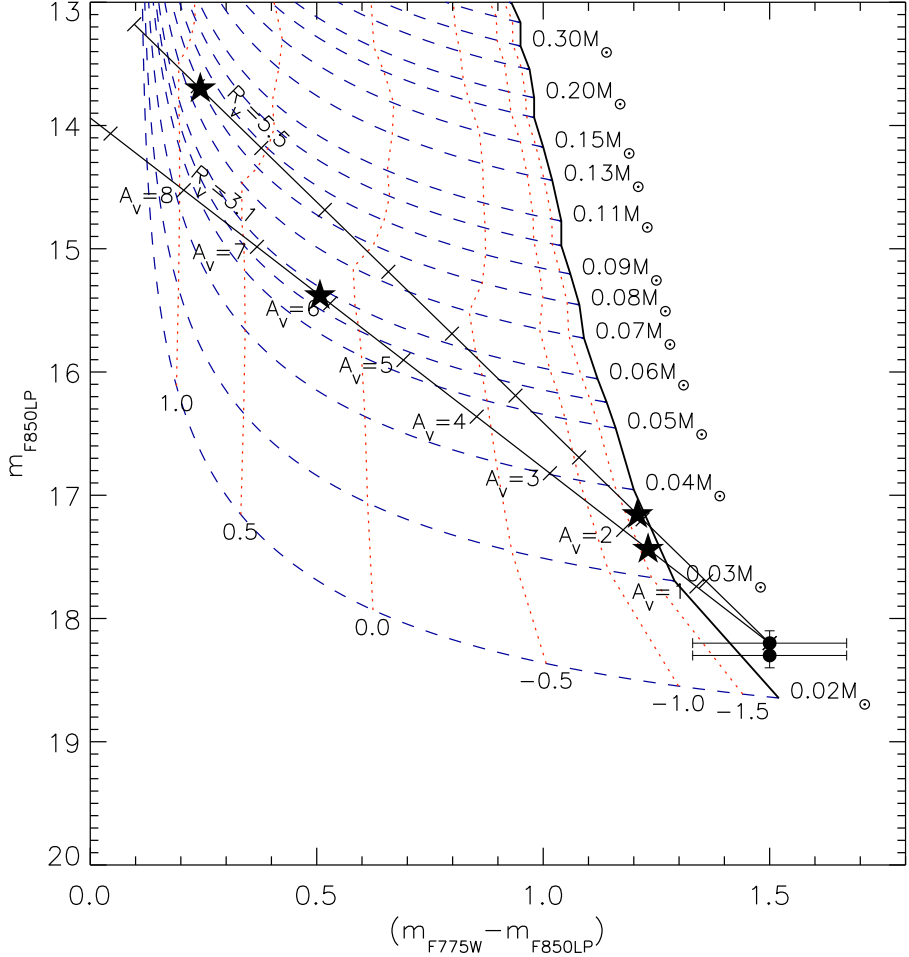}
 \caption{Color-magnitude diagram for the 2 point-like sources (filled circles). The solid curve represents the \cite{1998A&A...337..403B} pre-main-sequence 1~Myr isochrone, with the relative masses labeled along the curve. Blue dashed lines represent the displacement, for a star of a given mass, caused by mass accretion for increasing $\log(L_\mathrm{accr}/L_\mathrm{phot})$ ratio between accretion and photospheric luminosity. The red dotted lines join the
points of equal $\log (L_\mathrm{accr}/L_\mathrm{phot})$ values, labeled under the 0.02~$M_\sun$ dashed curve.  Dereddening vectors corresponding to extinction values $A_V=1,2,...,7$ are shown for two choices of the reddening
parameter $R_V$. The stars represent the solutions for the position of Source~N compatible with the color-color diagram shown in Figure~\ref{fig:color_color} (see text).  \label{fig:isochrone}}
\end{figure}

\begin{figure}
\epsscale{.80}
\plotone{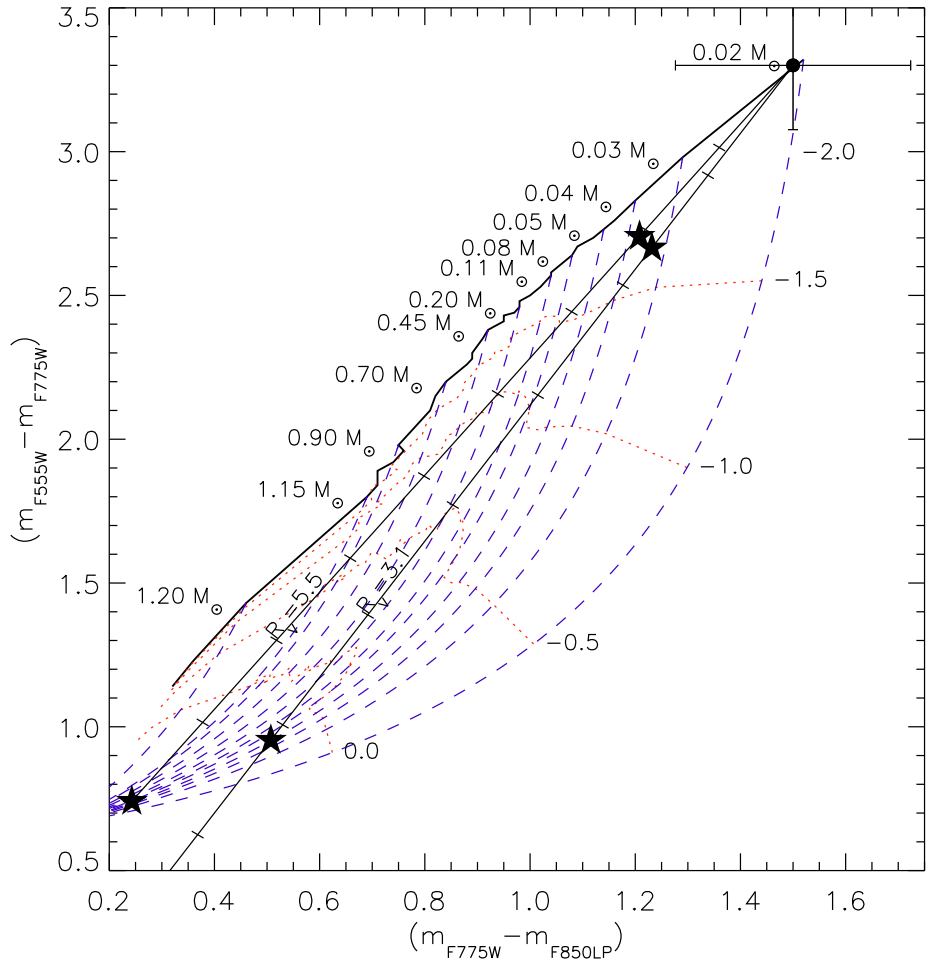}
\caption{Color-color diagram for Source N. 
Symbols are the same used in Figure~\ref{fig:isochrone}.  
\label{fig:color_color}}
\end{figure}

To constrain the nature of the two sources, we plot their position in a color-magnitude diagram (Fig.~\ref{fig:isochrone}) and (for Source~N only) in a color-color diagram (Fig.~\ref{fig:color_color}) with respect to the 1~Myr PMS isochrone of \cite{1998A&A...337..403B}, calculated in our passbands using the \textsc{NextGenII} synthetic atmosphere models \citep{nextgenII}. We show the
stellar population of the ONC, we also show the displacement from the isochrone caused by extinction and mass accretion, both combined with the synthetic stellar spectra before convolution with the filter passbands. For the extinction we use both the standard $R_V=3.1$ reddening law and the $R_V=5.5$ reddening law, considered more appropriate for the Orion Nebula 
\citep{BaadeMinkowsky37, Johnson67}. The accretion spectrum is reproduced following the \cite{CalvetGullbring98} recipe,  combining $3/4$ of a $T_\mathrm{eff}=7000$~K black body, accounting for the optically thick emission from the photospheric region heated by the shock of accreting material, with $1/4$ of optically thin emission from a dense slab ($n=10^8$~cm$^{-3}$) of photoionized gas also at $T_\mathrm{eff}\simeq7000$~K, accounting for the pre-shock region. The  accretion luminosity, $L_\mathrm{accr}$, and the stellar bolometric luminosity, $L_\mathrm{phot}$, are both calculated integrating the model spectra, with $L_\mathrm{accr}$ finally multiplied by a normalizing factor. This approximate treatment is adequate for our purposes, as we are interested in the intensity in broad-band filters rather than in the details of the true accretion spectrum.

\begin{figure}
\epsscale{1.16}
\plottwo{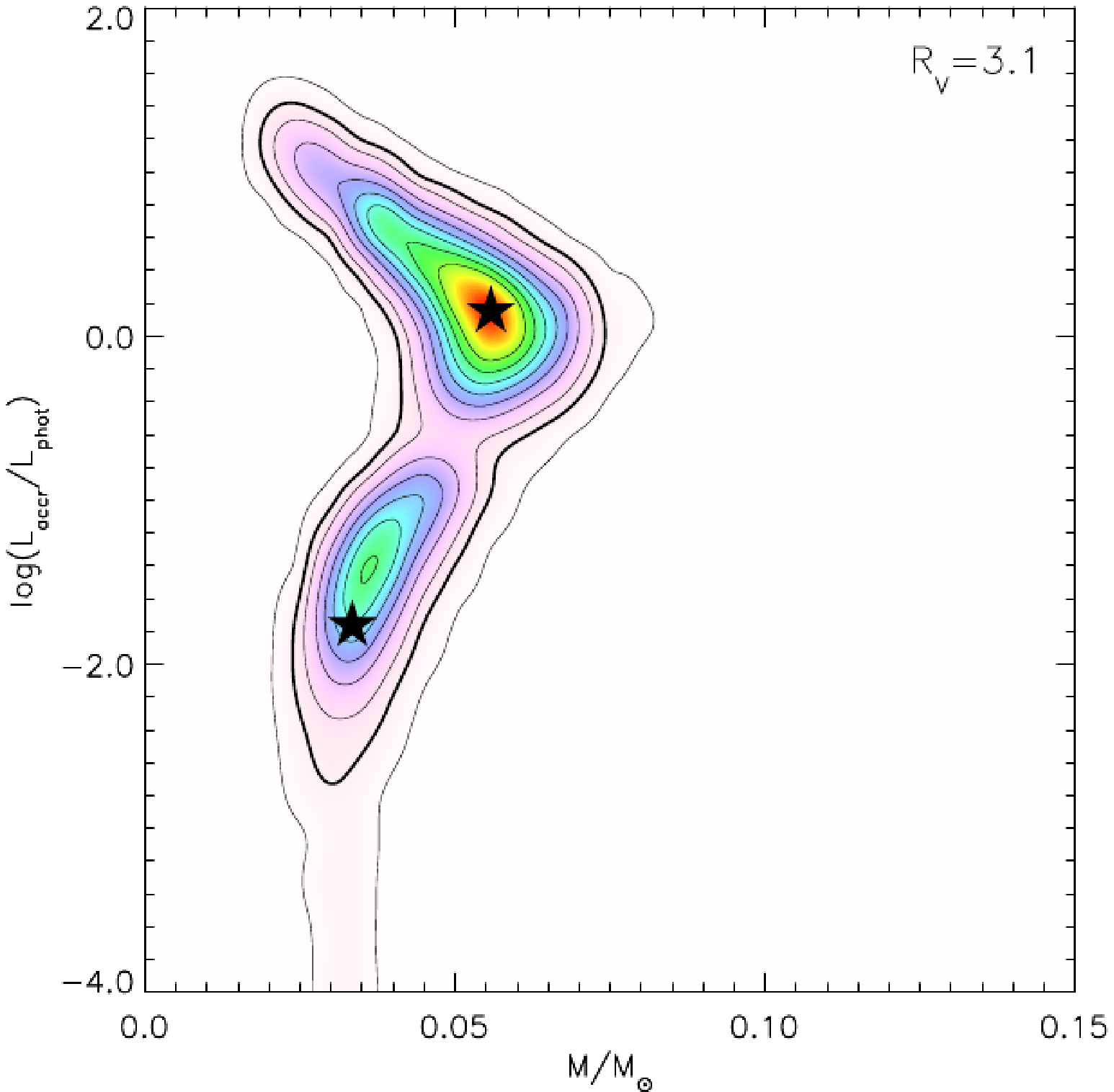}{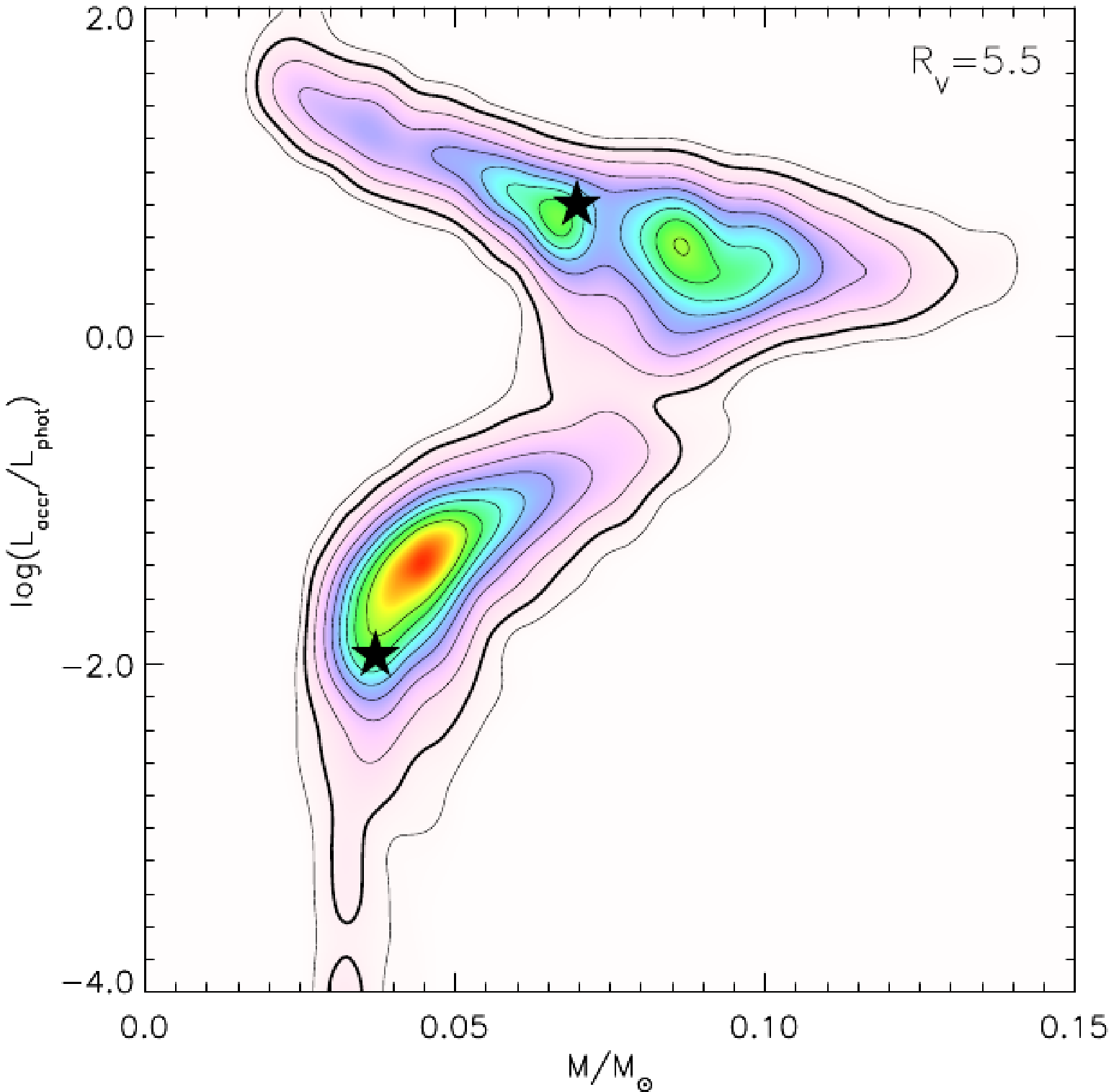}
\caption{Distribution of mass and accretion luminosity for Source~N, derived from our Monte Carlo simulation. The stars refer to the nominal position of the source (see text). The left and right panel are relative to the $R_V=3.1$ and $R_V=5.5$ reddening law, respectively. \label{fig:mass_accr_relation}}
\end{figure}

Figures~\ref{fig:isochrone} and \ref{fig:color_color} show that the observed  magnitudes of Source~N can be reproduced by different combinations of $A_V$, $\log(L_\mathrm{accr}/L_\mathrm{phot})$ and stellar mass. To find the best combinations of parameters, we calculate a large 3-d grid of models covering the  full range of viable $A_V$, $\log(L_\mathrm{accr}/L_\mathrm{phot})$ and stellar mass, derive the synthetic photometry and search for the best match with the data.  The results are summarized in Table~\ref{table:intersections} and  marked as stars in Figures~\ref{fig:isochrone} and \ref{fig:color_color}. It turns out that for each reddening law, there are typically two intersection points. For each reddening law, the two intersections provide combinations of lower and higher mass, extinction and accretion, (the \textit{low} and \textit{high} solutions). All four solutions place Source~N in the brown dwarf regime. To test the robustness of this result and to find the optimal solution, we have assumed that the uncertainties are dominated by photometric errors and performed a Monte Carlo simulation. We generated a set of 10,000 trial stars populating a gaussian error distribution for the three magnitudes; we then iterated our procedure to find the corresponding  mass, accretion luminosity and extinction. Figure~\ref{fig:mass_accr_relation} shows the results for the two choices of $R_V$. The two areas at the top and bottom of each figure  correspond to the \textit{high} and \textit{low} solutions, respectively. The  height of the peaks shows  that the \textit{high} solution is more compatible with $R_V=3.1$, vice versa if $R_V=5.5$. Only if  $R_V=5.5$ the strongly degenerate \textit{high} solution may lead to masses above the hydrogen burning limit ($M\simeq0.08~M_\odot$).  

Another clue to the nature of the sources may come from a comparison of our estimated mass accretion rate with the values normally found for similar stars. Using Eq.~(2) of \cite{Muzerolle+03} with the stellar parameters from our Baraffe/NEXTGEN models, we derive for $\dot{M}$ the values reported in Table~\ref{table:intersections}. A comparison with Figure~8 of \cite{Muzerolle+03}, or Figure~2 of \cite{Mohanty+05}, shows that only our \textit{low} solutions provide mass accretion rates similar to those found in low-mass BDs, whereas the \textit{high} solutions give values in excess by 2~orders of magnitudes. If we exclude the \textit{high} values on this basis,  the most consistent solution  for Source~N is the \textit{low} case, corresponding to a $M\simeq 0.04~M_\odot$ brown-dwarf accreting at a rather standard $\dot{M}\simeq 5\times10^{-11}~M_\odot$~yr$^{-1}$, with about $A_V=2$ and $R_V=5.5$, indicative of grains larger than those typically found in the interstellar medium. Similar conclusions should hold for Source~S, given the nearly identical $I$ and $z$~magnitudes. Even if IR spectroscopy will be needed to firmly establish the nature of these objects, the possibility that the two sources have sub-stellar mass appears quite robust. If confirmed, this would be the first direct observation of a young brown-dwarf binary still accreting in its circumbinary disk.

Whereas our images shows the presence of a large-scale circumbinary disk, the evidence for mass accretion points to to the presence of at least one circumstellar disk. Theory provides a consistent scenario, predicting that circumstellar and circumbinary disks emerge as the mature outcome of the evolution of the original circumstellar disk in which the binary system formed
\citep{Gunther+02}. Due to the exchange of angular momentum from the binary to the disk, an inner gap develops between the inner and outer disk regions. Material can flow through the gap along spiral arms, feeding the circumstellar disks and therefore sustaining mass accretion into the central stars. A detailed comparison of our system with the extant theoretical models is hindered, however, by the fact that our circumbinary disk is being photoevaporated, the first time such a phenomenon is observed. The influence of UV flux from an external source on a circumstellar disk has been studied only for single stars by  \cite{2002ApJ...578..897R}, who have shown that the extra heating of a disk face produces an increase of the disk flaring angle, leading to photoevaporation and possibly a warping of the system. This may explain why Source~S remains  undetected in our bluest filters:  the southern side of the disk is the one more directly exposed to the UV flux and the stronger photoevaporation may locally increase the optical depth on this side. Assuming the sources are identical, a line-of-sight difference $\Delta A_V\simeq 0.3^{m}$  would bring Source~S below our detection limit at the shortest wavelengths while being compatible with the magnitudes observed in the red filters.
\vspace{-12pt}
\acknowledgements
The work of LR and ND at STScI was done under the auspices of the STScI  Summer Student Program. The authors wish to thank the anonymous referee for prompt and extremely valuable comments. 

{\it Facilities:} \facility{HST (ACS/WFC)}.

\clearpage

\begin{deluxetable}{ccc}
\tablecaption{Source Photometry\label{tbl-2}}
\tablewidth{0pt}
\tablehead{
\colhead{Filter} & \colhead{Source N} & \colhead{Source S}
}
\startdata
F435W & 23.3 & --\\
F555W & 23.0 & 23.5\\
F658N & 19.2 & -- \\
F775W & 19.7 &19.8\\
F850LP & 18.2 & 18.3\\
\enddata
\tablecomments{Magnitudes in Vega photometric system.}
\end{deluxetable}

\begin{deluxetable}{cccccc}
\tablewidth{0pt}
\tablecaption{Extinction, mass, accretion luminosity and mass accretion rate for Source N.  \label{table:intersections}}
\tablehead{
\colhead{case}&
\colhead{$R_V$}&
\colhead{$A_V$}&
\colhead{$M$} &
\colhead{$\log(L_{\rm accr}/L_{\rm phot})$} &  
\colhead{$\log\dot{M}$}\\  
& & & ($M_{\odot}$) &\ & {($M_\sun$ yr$^{-1}$)}
}
\startdata
\textit{low} & $3.1$ & $1.66$ & $0.033$ & \llap{$-$}1.76 &$ -10.17$ \\
  & $5.5$ & $2.08$ & $0.037$ & \llap{$-$}1.95 & $-10.26$ \\
\textit{high} & $3.1$ & $6.14$ & $0.056$ &  $0.15$ &\phn$-7.83$ \\
  & $5.5$ & $8.96$ & $0.070$ &  $0.80$ &\phn$-7.02$  \\
\enddata
\end{deluxetable}

\clearpage 

\begin{thebibliography}{}

\bibitem[Artymowicz et~al.(1991)]{1991ApJ...370L..35A} Artymowicz, P., 
Clarke, C.~J., Lubow, S.~H., \& Pringle, J.~E.\ 1991, \apjl, 370, L35 
\bibitem[Artymowicz \& Lubow(1994)]{1994ApJ...421..651A} Artymowicz, P., \& Lubow, S.~H.\ 1994, \apj, 421, 651
\bibitem[Baade \& Minkowski(1937)]{BaadeMinkowsky37}Baade, W., \& Minkowski, R. 1937, \apj, 86, 123
\bibitem[Baraffe et~al.(1998)]{1998A&A...337..403B} Baraffe, I., Chabrier, G., Allard, F., \& Hauschildt, P.~H.\ 1998, \aap, 337, 403 
\bibitem[Bate \& Bonnell(1997)]{1997MNRAS.285...33B} Bate, M.~R., \& Bonnell, I.~A.\ 1997, \mnras, 285, 33 
\bibitem[Calvet \& Gullbring(1998)]{CalvetGullbring98} Calvet, N. \&\ Gullbring, E.\ 1998, \apj, 509, 802
\bibitem[Duchene(1999)]{Duchene+99} Duch{\^e}ne, G.\ 1999, \aap, 341, 547
\bibitem[Dutrey et~al.(1994)]{1994A&A...286..149D} Dutrey, A., Guilloteau, S., \& Simon, M.\ 1994, \aap, 286, 149 
\bibitem[Duvert et~al.(1998)]{1998A&A...332..867D} Duvert, G., Dutrey, A., Guilloteau, S., Menard, F., Schuster, K., Prato, L., \& Simon, M.\ 1998, \aap, 332, 867 
\bibitem[G{\"u}nther \& Kley(2002)]{Gunther+02} G{\"u}nther, R., \& Kley, W.\ 2002, \aap, 387, 550
\bibitem[Hauschildt et~al.(1999)]{nextgenII} Hauschildt, P.~H., Allard, F., Baron, E.\ 1999, \apj, 512, 377
\bibitem[Johnson(1967)]{Johnson67}Johnson, H. L. 1967, \apj, 150, L39
\bibitem[K\"ohler et~al.(2006)]{Koehler+06}
K\"ohler, R., Petr-Gotzens, M. G., McCaughrean, M. J., Bouvier, J., Duch\^ene, G., Quirrenbach, A., \&\ Zinnnecker, H. 2006, \aap, 458, 461
\bibitem[Lin \& Papaloizou(1993)]{1993prpl.conf..749L} Lin, D.~N.~C., \& Papaloizou, J.~C.~B.\ 1993, Protostars and Planets III, 749 
\bibitem[Luhman(2004)]{Luhman04}Luhman, K.~L.\ 2004, \apj, 614, 398 
\bibitem[Luhman et~al.(2007)]{Luhman+07}
Luhman, K. L., et al.  2007, \apj, 666, 1219
\bibitem[Mathieu(2000)]{Mathieu+00} Mathieu, R.~D.\ 2000, ``Pre-main-sequence Binary Stars" in Encyclopedia of Astronomy and Astrophysics, Edited by Paul Murdin, article 2384. Bristol: Institute of Physics Publishing, 2001
\bibitem[Menten et~al.(2007)]{2007A&A...474..515M} Menten, K.~M., Reid, M.~J., Forbrich, J., \& Brunthaler, A.\ 2007, \aap, 474, 515 
\bibitem[Mohanty et~al.(2005)]{Mohanty+05} Mohanty, S., Basri, G., \&\ Jayawardhana R. 2005, Astron.\ Nachr. 326, 891
\bibitem[Monin et~al.(2007)]{Monin+07} Monin, J.-L., Clarke,
C.~J., Prato, L., \& McCabe, C.\ 2007, Protostars and Planets V, 395
\bibitem[Muzerolle et~al.(2003)]{Muzerolle+03}Muzerolle, J., Hillenbrand, L., Calvet, N., Brice\~{n}o, C., Hartmann, L. 2003, \apj, 592, 266
\bibitem[Padgett et~al.(1997)]{Padgett+97} Padgett, D.~L., Strom,
S.~E., \& Ghez, A.\ 1997, \apj, 477, 705
\bibitem[Petr et~al.(1998)]{1998ApJ...500..825P} Petr, M.~G., Coude Du 
Foresto, V., Beckwith, S.~V.~W., Richichi, A., 
\& McCaughrean, M.~J.\ 1998, \apj, 500, 825 
\bibitem[Prosser et~al.(1994)]{Prosser+94} Prosser, C.~F.,
Stauffer, J.~R., Hartmann, L., Soderblom, D.~R., Jones, B.~F., Werner,
M.~W., \& McCaughrean, M.~J.\ 1994, \apj, 421, 517
\bibitem[Reipurth et~al.(2007)]{Reipurth+07}
Reipurth, B., Guimara\~es, M. M., Connelley, M. S., \&\ Bally, J. 2007, \aj, 134, 2272
\bibitem[Ricci et~al.(2008)]{Ricci+08}
Ricci, L., Robberto, M., \&\ Soderblom, D. R., 2008, \aj,  in press
\bibitem[Robberto et~al.(2004)]{2004ApJ...606..952R} Robberto, M., Song, J., Mora Carrillo, G., Beckwith, S.~V.~W., Makidon, R.~B., 
\& Panagia, N.\ 2004, \apj, 606, 952 
\bibitem[Robberto et~al.(2002)]{2002ApJ...578..897R} Robberto, M., Beckwith, S.~V.~W., \& Panagia, N.\ 2002, \apj, 578, 897 
\bibitem[Sirianni et~al.(2005)]{Sirianni+05} Sirianni, M., et al.\
2005, \pasp, 117, 1049
\bibitem[Smith et~al.(2005)]{Smith+05} Smith, N., Bally, J.,
Licht, D., \& Walawender, J.\ 2005, \aj, 129, 382
\end{thebibliography}
\end{document}